\documentclass[aps,pre,twocolumn,groupedaddress,showpacs]{revtex4-1}

\usepackage[pdftex]{graphicx}
\usepackage{enumerate}

\usepackage{amssymb,amsfonts,amsmath}

\begin{document} 

\title{Boundary Singularities Produced by the Motion of Soap Films}

\author{Raymond E Goldstein, James McTavish, H Keith Moffatt, and 
Adriana I Pesci}
\affiliation{Department of Applied Mathematics and Theoretical Physics, University of Cambridge, Wilberforce Road, Cambridge CB3 0WA, United Kingdom}

\begin{abstract}
Recent work has shown that a M{\"o}bius strip soap film rendered unstable by deforming its frame changes topology to that of a disk through 
a `neck-pinching' boundary singularity. This behavior is unlike that of the catenoid, which transitions to two disks through a bulk 
singularity. It is not yet understood whether the type of singularity is generally a consequence of the surface topology, nor how this 
dependence could arise from an equation of motion for the surface. To address these questions we investigate experimentally, computationally, 
and theoretically the route to singularities of soap films with different topologies, including a family of punctured Klein bottles. We 
show that the location of singularities (bulk or boundary) may depend on the path of the boundary deformation. In the unstable regime 
the driving force for soap-film motion is the mean curvature.  Thus, the narrowest part of the neck, associated with the shortest 
nontrivial closed geodesic of the surface, has the highest curvature and is the fastest-moving. Just before onset of the instability 
there exists on the stable surface a shortest closed geodesic, which is the initial condition for evolution of the neck's geodesics, 
all of which have the same topological relationship to the frame. We make the plausible conjectures that if the initial geodesic is 
linked to the boundary then the singularity will occur at the boundary, whereas if the two are
unlinked initially then the singularity will occur in the bulk. Numerical study of mean curvature flows and experiments support 
these conjectures.
\end{abstract}


\maketitle

\noindent {\it Significance Statement:
The dynamics of topological rearrangements of surfaces are generically associated with singularities.  It is not yet understood whether the 
location of a singularity (in the bulk or at boundaries) is a general consequence of the surface topology, nor how this dependence could 
arise from an equation of motion for the surface.  Here we use experiments on soap films, computation,  and theory to introduce simple and 
intuitive concepts to describe generic transitions and their singularities. In particular, we propose a criterion to predict, from the 
configuration of the initial state, the location of the singularity.  This criterion involves the topological linkage between the bounding 
curve of the surface and a local `systole', the local non-trivial minimum-length closed curve on the surface.
}

\vskip 0.5cm

The study of singularities, be it their dynamic evolution \cite{Heraklion, Tubes, Eggers} or their location and structure in
a  system at equilibrium \cite{Mermin, Arnold}, has a long history.  The main focus of research has
been either on systems without boundaries,  where  both the  static and dynamic cases are fairly 
well understood, or on finite systems with prescribed boundaries displaying static singularities, as for example
 defects in liquid crystals \cite{Langer} and superfluids \cite{Kasamatsu},  Langmuir monolayers \cite{LangmuirRef}, 
and quantum field theories \cite{Tong}. Work on dynamic aspects of these 
bounded systems is limited, particularly in the case of singularity formation accompanying a topological 
change \cite{Nonlinearity}. For instance, to our knowledge, there is no mathematical tool available to predict from initial conditions
whether a singularity in a bounded system will occur in the bulk or at the boundary.

In a previous paper \cite{GMPR} we found that a soap film in the shape of a  M{\"o}bius band spanning a slowly deforming 
wire would change its topology through a singularity that occurs at the boundary. This simple example 
provided a first model to investigate the dynamical aspects of the formation of a
boundary singularity from the moment the system becomes unstable, and until a new stable 
configuration is reached. While progress was made in the particular case of the M{\"o}bius band, left
unanswered were questions of greater generality: (i) Are there any
other configurations of films spanning a frame that have a topology transition with a boundary singularity, showing that
this behavior is generic, rather than exceptional? (ii) If so: What are  the possible types of evolution 
equations that could describe the dynamics of the collapse? (iii) When such a system becomes unstable
can topological and geometric parameters at onset be used to predict whether the 
singularity  will occur in the bulk or at the boundary?
As described below, we have found that the first of the questions has an affirmative answer. 
Just as Courant \cite{Courant1, Courant} used soap films as a means of studying minimal surface topology, so we
can seek to answer these questions through study of soap-film transition singularities.

The evolution of bulk singularities and of soap films moving towards equilibrium configurations have been extensively 
studied \cite{ColdingLong} with mean curvature flow, in which the surface moves normal to itself at a speed proportional
to the local mean curvature, and also with models more faithful to
hydrodynamics \cite{LeppinenLister,NitscheSteen}. There is also a large body of literature pertaining to the case of 
closed surfaces for which 
rigorous results have been proven \cite{Gage, GageHamilton, Grayson, Huisken}.  
In contrast, the  
dynamics of a surface with a deformable boundary which, after becoming unstable, undergoes a singular topological change
has received little or no attention, particularly so when the singularity occurs at the boundary.

\begin{figure*}[t]
\centering
\includegraphics[width=1.85\columnwidth]{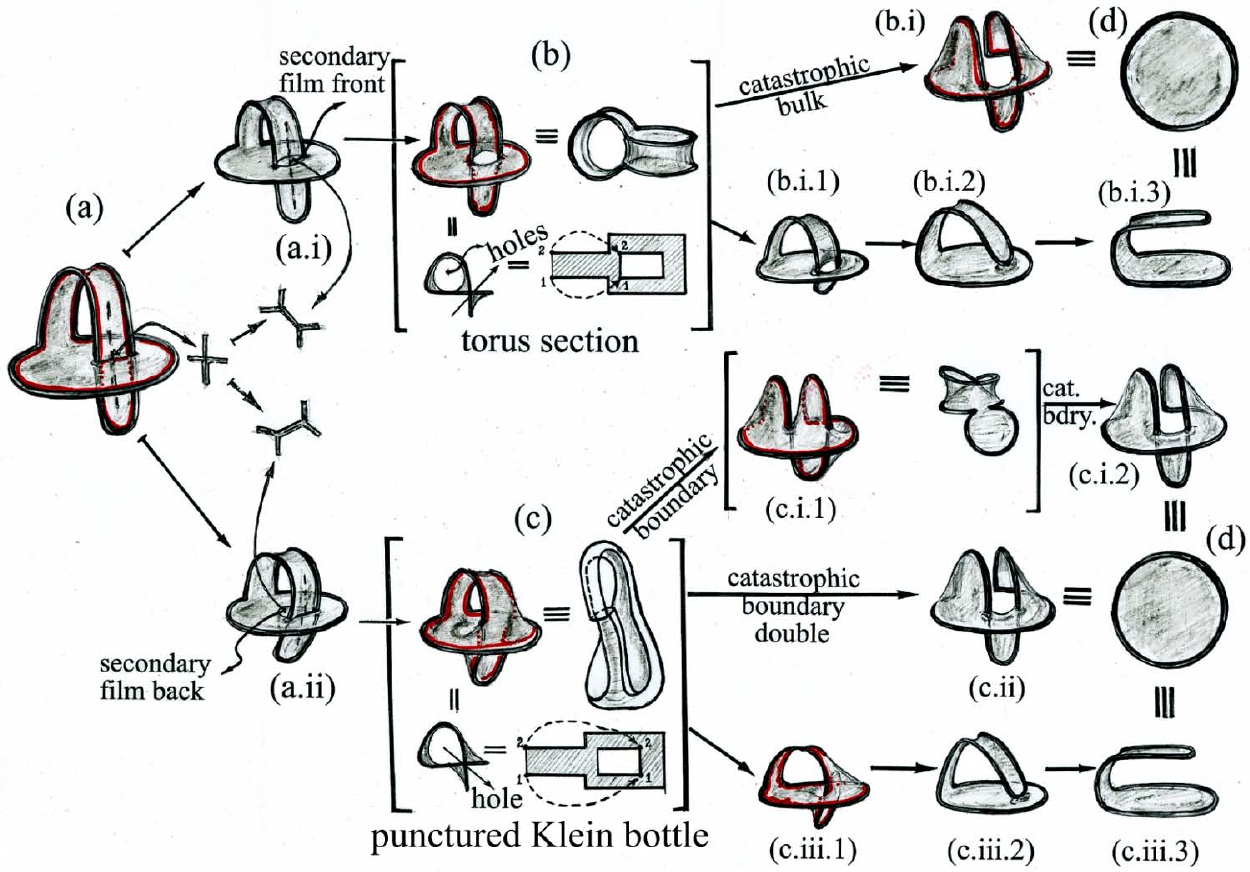}
\caption{Interconversions of various soap film topologies, as described in text. Shaded regions are films, and red lines 
denote locations of Plateau borders. \label{fig1}}
\end{figure*}

To study the dynamical aspects of boundary singularities, and their relationship to the topological and
geometric parameters of the film, 
we performed numerical calculations based on mean curvature flow and verified that, even though the actual
motion of a film also involves inertial and more realistic viscous forces, this approach is capable of reproducing  the 
essential characteristics of the boundary singularities we observed in experiment.  This stands in clear contrast with the case
of bulk singularities, where the mean curvature approach fails to reproduce fully the experimental observations.
For example, a film collapsing in the bulk by mean curvature produces
one singularity, whereas in experiment there are at least two placed above and below a satellite drop 
\cite{CryerSteen,ChenSteen97,RobinsonSteen}.

The best known example of such a bulk singularity in a bounded soap film is the collapse of a catenoid 
spanning two parallel rings. In this case the singularity occurs only when the rings have been separated beyond a critical 
distance and the film, as it becomes unstable, develops a narrow neck  on which a non-contractible curve of least
length, or `systole relative to the boundary' (SRB) \cite{Gromov,Guth_overview,Berger}, can be found.
Even though the collapse of the catenoid is a situation characterized by high symmetry, we will see here, 
through a combination of experiment and computation, that there is a whole 
class of singularities that follow the same pattern of neck formation such that 
infinitesimally close to but before the onset of the instability it is possible to identify a `local' SRB (LSRB)
that constitutes the initial condition for the evolution of the curve whose contraction 
ends at the location of the singularity (see Fig. \ref{fig5}).  We find that this feature of the initial LSRB and its linkage to the
boundary makes it possible to deduce whether the singularity occurs in the bulk or at the boundary, and also that 
the way in which the same boundary is deformed before onset of the instability can produce different types
of singularities.

\section{Results and Discussion}

\subsection{1. Experiments} In previous work  \cite{GMPR} it  was found that when a soap film in the shape of
a  M{\"o}bius band becomes unstable it collides with its frame via a neck-like singularity. This collision
yields changes in three topological quantities: the genus, the sidedness, and the  linking number between the Plateau 
border and the frame (for details, see \cite{GMPR}). These experiments could not, by themselves, establish if there was causality in these results, so we sought other
film configurations to test which of the topological parameters, if any, are predictors of the transition. One such configuration 
starts from the surface first introduced by Almgren \cite{Almgren} (see Fig. \ref{fig1}a). 
This film has the interesting feature of having an intersection along a segment that can develop into a secondary film by slightly pushing the
descending tab. The direction in which the tab is pushed determines if the secondary film lies towards
the front or the back (Figs. \ref{fig1}a.i and a.ii, respectively).  As illustrated, the two possible resolutions of the 
unstable four-fold vertex into the films in Fig. \ref{fig1}a.i and a.ii correspond to the two stable states which, in the
language of foams, interconvert through a T1 process \cite{foams}.
Puncturing the secondary film when it lies towards the front 
produces the well-known result 
\cite{Almgren} shown in Fig. \ref{fig1}b, which is topologically equivalent to a section of a torus.  
On the other hand, puncturing the secondary film when it lies towards the back produces a 
surface (Fig. \ref{fig1}c)  that we believe has not been identified before in this context, and which corresponds to a 
punctured Klein 
bottle, i.e. two M{\"o}bius bands of opposite chirality sewn together over part of their boundaries.  It is this latter 
surface that we used 
to test our hypotheses. This punctured Klein bottle ({\it punctured Klein}) is an ideal choice because,
like the M{\"o}bius band (linking number 2, non-orientable genus 1), it is one-sided, but its Plateau border linking number with 
the frame is zero and its non-orientable genus is 2, allowing us to separate the two topological parameters.

By pulling apart the top sections of the frame in Fig. \ref{fig1}c the punctured Klein becomes unstable and undergoes a transition 
to a disk by means of two mirror image singularities. This 
transition, which is identical to two simultaneous M{\"o}bius transitions of opposite chirality, becomes fully dynamical after 
onset and once it has finished (Fig. \ref{fig1}c.ii, or equivalently Fig. \ref{fig1}c.i.2) the parameters that have changed are the 
sidedness and the genus. There are at least two other alternative routes from the punctured Klein to the disk. 
The second route, although much harder to 
realize experimentally, is very intuitive mathematically: since the punctured Klein is made of two sewn M{\"o}bius bands
it is possible to first coerce the frame into collapsing one of the bands producing a change of non-orientable genus from 2 to 1 and also 
from linking number from 0 to $\pm 2$, and leaving as a result only one of the M{\"o}bius bands of the original 
pair (Fig. \ref{fig1}c.i.1). Hence  the surface is still one-sided. The last part of the process is the usual M{\"o}bius collapse that will yield the
two-sided disk (Fig. \ref{fig1}c.i.2) and will again change both the genus and the  linking number to zero, as expected. 
In the end the features of the collapse to a final two-sided disk can be distilled to those of the two individual M{\"o}bius bands that made the whole
and as a consequence each one of these two steps is a fully dynamical process.
The fact that any other film that can be constructed in such a way that it consists of a series of M{\"o}bius bands sewn together can be 
analyzed in a similar manner hints to the possibility of constructing a simple algebra to keep track of the net changes of
the topological parameters of the surface at every stage of the collapse.   The  third way to convert the punctured 
Klein into a disk
requires deforming the boundary quasi-statically by pulling the tab upwards slowly until its tip rises above the plane of the 
horizontal film (Figs. \ref{fig1}c.iii.1 - c.iii.3). Along this route, unlike along the other two, the process can be 
stopped at any  instant of time and there
will always be a stable minimal surface that spans the frame. When the tip of the tab finally reaches the horizontal plane
(d) the twists of the Plateau borders and the `holes' of the film will all converge to a point (the location of the singularity). The
twists will annihilate each other because they are of opposite chirality and the hole will vanish to a point. In the physical film the 
hole would disappear by reconnecting. All these different routes to the end result share the feature that the singularity 
always occurs at the boundary.

In the same way in which a whole  class of transitions in the bulk reduces asymptotically to the neck-like transition of the catenoid, we 
expect that  a whole class of boundary transitions of one-sided surfaces  will reduce to the the neck-like transition of a M{\"o}bius band. Thus,
studying the collapses of these two surfaces should be sufficient to cover all possible 
types of neck-like singularities.

There are three possible types of collapse:
\begin{enumerate}[(i)]
\item {\it Quasi-static boundary singularity}: the surface changes its topology in such a way that a stable minimal surface
can be found at every instant before and up to the time of the singularity which 
occurs at the boundary.

\item {\it Catastrophic bulk singularity}:  the surface changes its topology  after becoming unstable. Hence, the process is fully
 dynamic  and there are no intermediate stable minimal surfaces  visited in between the last stable  surface before onset 
and until  the final stable surface  is reached. In this case the singularity is located in the bulk

\item {\it Catastrophic boundary singularity}: the same as (ii) except that the singularity occurs at the boundary. 
\end{enumerate}
While there is a fourth possibility -- a quasi-static bulk singularity -- we have not found a film configuration 
where this is observed. 
The different types of singularities have a specific correlation with the change of topology, but beyond the change 
of genus, there is no one-to-one correspondence between the type of transition and the topological parameters. The rules that govern
the transitions are:
\vskip -0.7cm
\begin{enumerate}[(a)]
\item In a catastrophic boundary singularity there is a change in either the linking number or sidedness, or both.

\item If a change of linking number or sidedness occurs there is a boundary singularity, either catastrophic or quasi-static.

\item A catastrophic bulk singularity can change neither the linking number nor the sidedness of the surface.

\item A quasi-static boundary singularity can support any type of topological change.
\end{enumerate}
\vskip -0.3cm
From our studies it appears that both types of catastrophic singularities, bulk and boundary, can be transformed into a 
boundary quasi-static singularity by choosing a suitable frame deformation. For example, for the catenoid, instead of pulling the 
rings apart to render the film unstable and force the transition to two disks we can tilt the top ring quasi-statically until it reaches a 
position where the film touches itself and the boundary, thus creating a pair of disks connected by a single point. The connection 
point is the location of the singularity that occurs as the two disks are finally separated.

\begin{figure}[t]
\centering
\includegraphics[width=1.0\columnwidth]{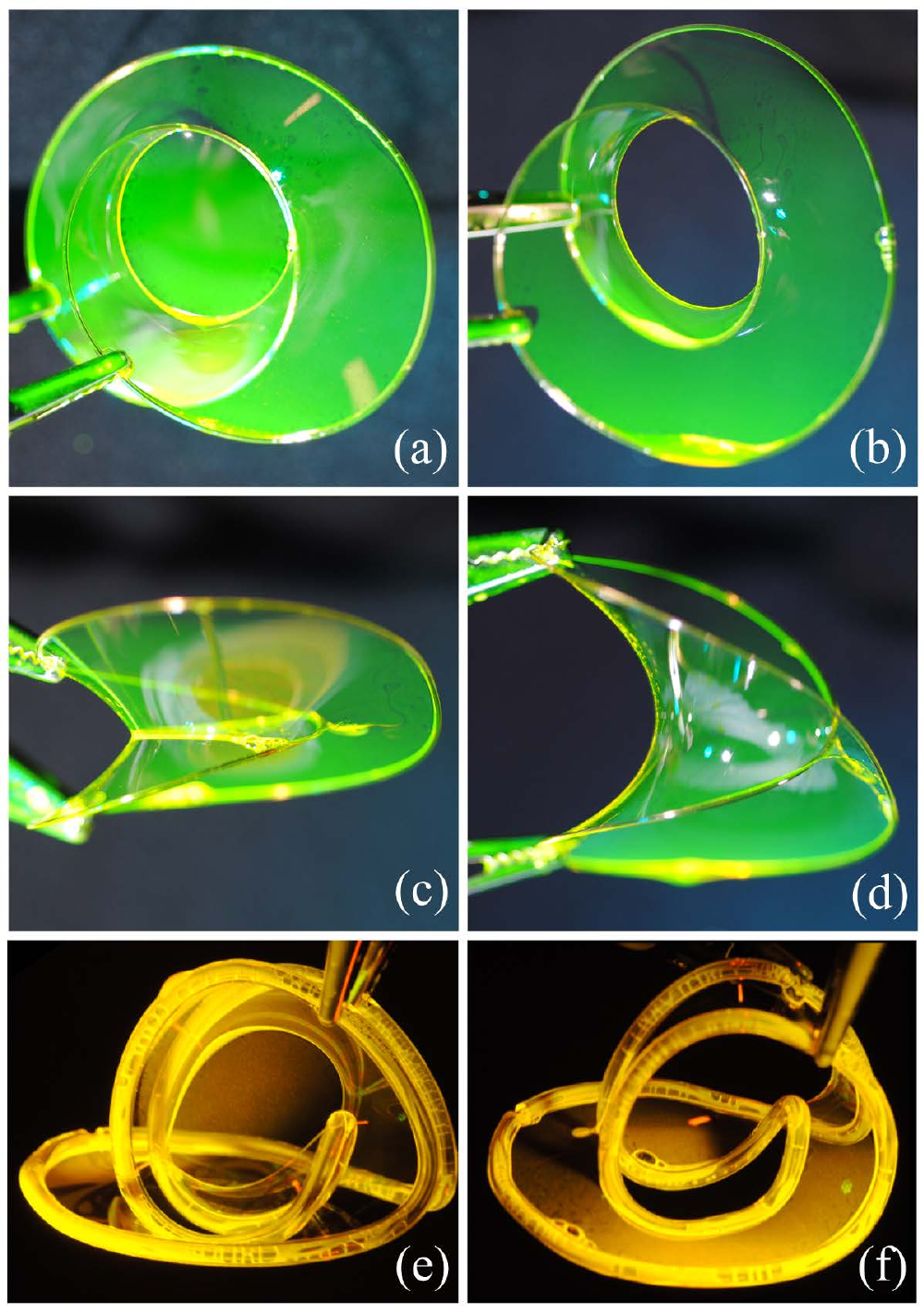}
\caption{Soap films. Top (a, b) and side (c, d) views of M{\"o}bius surfaces with and without secondary film. 
(e, f) Punctured Klein bottled surfaces with and without secondary film.\label{fig2}}
\end{figure}

The experiments also helped with the study of another feature of these films  that is connected with geometric characteristics
of the associated minimal surfaces.  In general, when films that will collapse through a neck-like singularity are made, they come
out of the  soap solution with a secondary film spanning the neck (Figures \ref{fig2}a, c and e). 
This feature was highlighted by Almgren and collaborators \cite{Almgren},
who named the shared curve between the main and secondary films the singular set.  
If the secondary film is allowed to remain while the 
wire is deformed it is possible to contract it to a point so that the final surface is 
identical to the one that would have been obtained if the secondary film were
removed at the start of the process. Because soap films minimize area it is clear that the location of the singular set is also the 
location of the shortest closed non-contractible geodesic on the main film. In fact, we observe that secondary films generically 
occur in situations where 
the main film supports the existence of such a curve. When the secondary film contracts during a transition, this geodesic 
shrinks and eventually reduces to a point which coincides with the location of the singularity. Not surprisingly, in the 
absence of the secondary film, the shortest closed geodesic of the main film is close to the original singular set. In a 
physical film the shortest closed 
geodesic in the main film is, however, longer than the curve determined by the singular set due to the effect of surface tension.

\subsection{2. Theoretical and Numerical Results} In this section we first address whether motion by mean curvature 
is capable of reproducing the experimental observations of boundary
singularities in the collapse of M{\"o}bius band and punctured Klein surfaces. This requires mathematical representations of boundary  
curves for the two cases which contain parameters describing the deformations leading to surfaces instabilities. For the 
M{\"o}bius band we have previously identified \cite{GMPR} the one-parameter family of curves
${\rm C}_M\,:\,{\bf x}(\theta)=(x,y,z)$
with $\mu = -1$ and $0 \le \theta < 2\pi$, where \cite{Maggioni}
\begin{subequations}
\label{eqn1}
\begin{eqnarray}
x(\theta)&=&\frac{1}{\ell_M}[\mu\tau\cos\theta+(1-\tau)\cos 2\theta]~,\\
y(\theta)&=&\frac{1}{\ell_M}[\mu\tau\sin\theta+(1-\tau)\sin 2\theta],\\
z(\theta)&=&\frac{1}{\ell_M}[2\mu\tau(1-\tau)\sin\theta]~.
\end{eqnarray}
\end{subequations}
Here, the parameter $\tau$ ($0 \le\tau\le 1$) interpolates smoothly from the double-covering of the circle for $\tau=0$ to a single circle
in the $xy$-plane when $\tau=1$.  The factor $\ell_M(\tau)$ normalizes the wire length to $2\pi$.  Numerical studies with Surface
Evolver \cite{Brakke}
show that 
the one-sided minimal surface spanning this curve becomes unstable when $\tau>\tau_{1c} \simeq 0.398$,
whereas the two-sided surface becomes unstable for $\tau<\tau_{2c}\simeq 0.225$.  In between these values the system is bistable, with
the energies of the two surfaces crossing at $\tau\simeq 0.38$.

\begin{figure}[t]
\centering
\includegraphics[width=1.0\columnwidth]{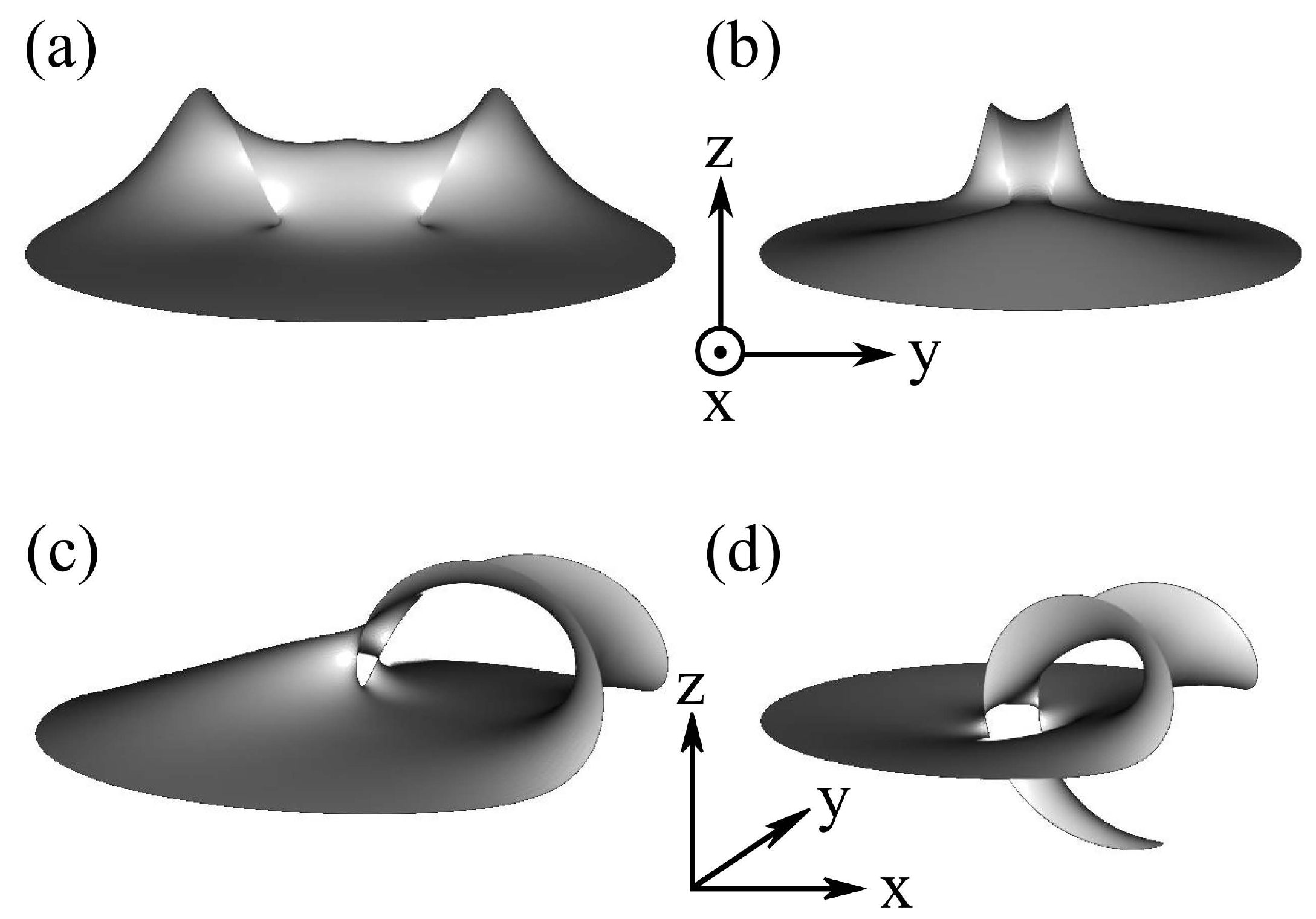}
\caption{Numerically obtained minimal punctured Klein surfaces. Top row: Varying the parameter $\lambda$ for $t=3.5$. (a) $\lambda=10$,
(b) $\lambda=120$.  Bottom row: Varying $t$ for $\lambda=10$. (c) $t=3$, (d) $t=5$. \label{fig3}}
\end{figure}

The frame shown in Fig. \ref{fig1}a, which can support the punctured Klein surface (\ref{fig1}c), can be approximated by a two-parameter
family of curves
${\rm C}_K\,:\,{\bf x}(\theta)=(x,y,z)$
for $0 \le \theta < 2\pi$, where
\begin{subequations}
\label{eqn2}
\begin{eqnarray}
x(\theta)&=&\frac{1}{\ell_K}\left\{\cos\theta+{1\over 2}\cos\left[ t\ {\rm e}^{-\lambda\sin^2\theta/2}\right]\right\}~,\\
y(\theta)&=&\frac{1}{\ell_K}\sin\theta~,\\
z(\theta)&=&\frac{1}{2(1+\sin^2\theta/2)\ell_K}~\sin\left[ t\ {\rm e}^{-\lambda\sin^2\theta/2}\right]~,
\end{eqnarray}
\end{subequations}
and $\ell_K(t,\lambda)$ is again a normalization, here depending on the two parameters of the curve.
As shown in Fig. \ref{fig3}, the parameter $t$ primarily controls the length of the `tab' below the plane of the circle, while
$\lambda$ controls the width of the tab.   

In each of these cases a dynamical evolution towards the singularity was obtained by starting with an equilibrium
shape close to the critical parameter(s) for instability and then incrementing one of the parameter(s) a small amount beyond the 
threshold. The subsequent evolution of the surface was calculated using the Hessian command in Surface Evolver with repeated
equiangulations and vertex averaging to ensure a well-behaved mesh.  
In both the M{\"o}bius and punctured Klein cases we find that the 
singularity or singularities are indeed located at the frame.   These dynamical evolutions are shown in Movies S1 \& S2 
(M{\"o}bius) and S3 \& S4 (punctured Klein).

\begin{figure}[t]
\centering
\includegraphics[width=1.0\columnwidth]{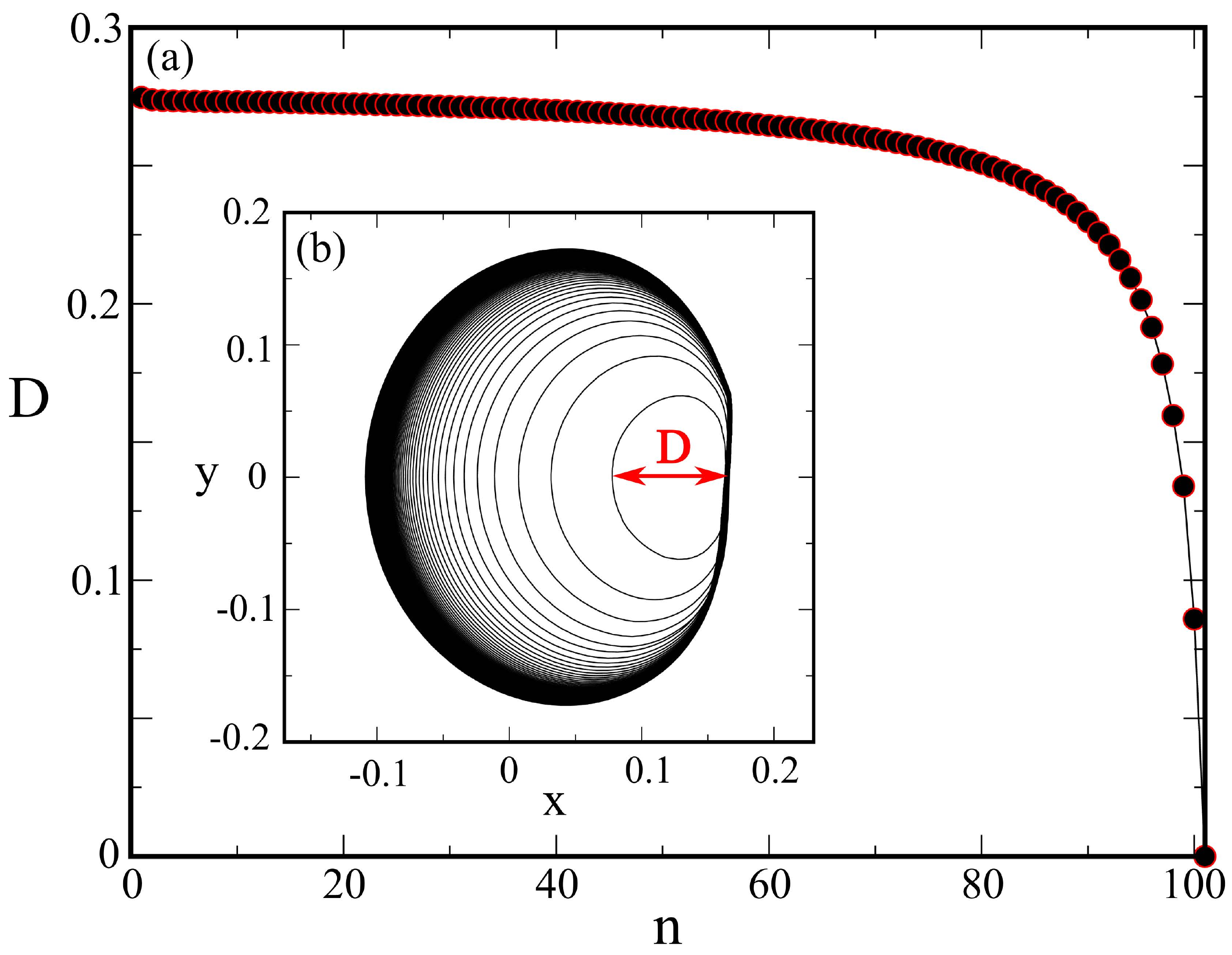}
\caption{Collapse dynamics of a M{\"o}bius surface from Surface Evolver.  (a) Diameter of the projection of geodesics 
(b) onto the $x-y$ plane (see Fig. \ref{fig3}) as
a function of iteration number $n$ in the relaxation scheme.  The singularity appears at the boundary at a finite time. \label{fig4}}
\end{figure}

To quantify the collapse of a neck to the boundary we determined from each triangulated surface the narrowest `diameter' 
$D$ of the neck 
measured from the point of the eventual location of the singularity to the point on the shortest non-contractible geodesic of the 
neck at which the normal is the  generating vector of the line that joins both points (Fig. \ref{fig4}).  That geodesic, in turn, 
was found by means of an 
adaptation of Kirsanov's Matlab program \cite{Kirsanov} to find exact geodesics between two specified endpoints on a triangular mesh.  
To find {\it closed} geodesics we implemented an iterative procedure to adjust the positions of
three endpoints spaced around the neck until the three geodesic segments formed a single smooth curve.
Figure \ref{fig4}  shows the projections of the geodesics onto the $x-y$ plane and the neck diameter $D$
for the M{\"o}bius collapse as a function of iteration number of the relaxation 
scheme.  Observe that the stationary right-hand section of the curves in Fig. \ref{fig4}b are the sections that the geodesics share
with the boundary.  The results indicate a clear collapse to the boundary in finite time.  While the data appear consistent with a
power-law variation of $D$ with time, with a sublinear exponent, at present there is insufficient resolution to determine that exponent 
more precisely.

As we have mentioned before, motion by mean curvature cannot faithfully represent a bulk singularity in a physical 
film because in such a film the effects of air pressure have a crucial role in the last stages of the collapse when a 
Rayleigh instability is triggered and  the film produces 
a satellite drop (or cascade of drops with a cut off controlled  by the parameters of the film) \cite{Eggers,ChenSteen97}. 
The difference between this singularity 
and the boundary one that makes  the latter amenable to description  with mean curvature flow is that physical effects other than
surface tension, for instance thickness variations in the film, become relevant only after the singularity has fully developed. In fact, the reconnection of the Plateau border in the collision with the boundary occurs 
{\it after} the surface has reached the location of the singularity.

{\it Location of singularity.} Systoles, or shortest non-contractible geodesics, have been extensively studied 
in the context of general
closed orientable Riemannian manifolds \cite{Gromov}.   In fact, inequalities relating the systole's length to the surface area and volume of the
manifold are well known.
Each one of the shortest non-contractible closed geodesics that we used to characterize the neck evolution under
mean curvature flow is an example of a local systole with respect to the boundary (LSRB), as defined in the introduction.
While up until now we have only found each local systole as the neck shrinks, it is also possible to find the one such curve
on the stable minimal surface prior to the onset of the instability.

\begin{figure}[t]
\centering
\includegraphics[width=1.0\columnwidth]{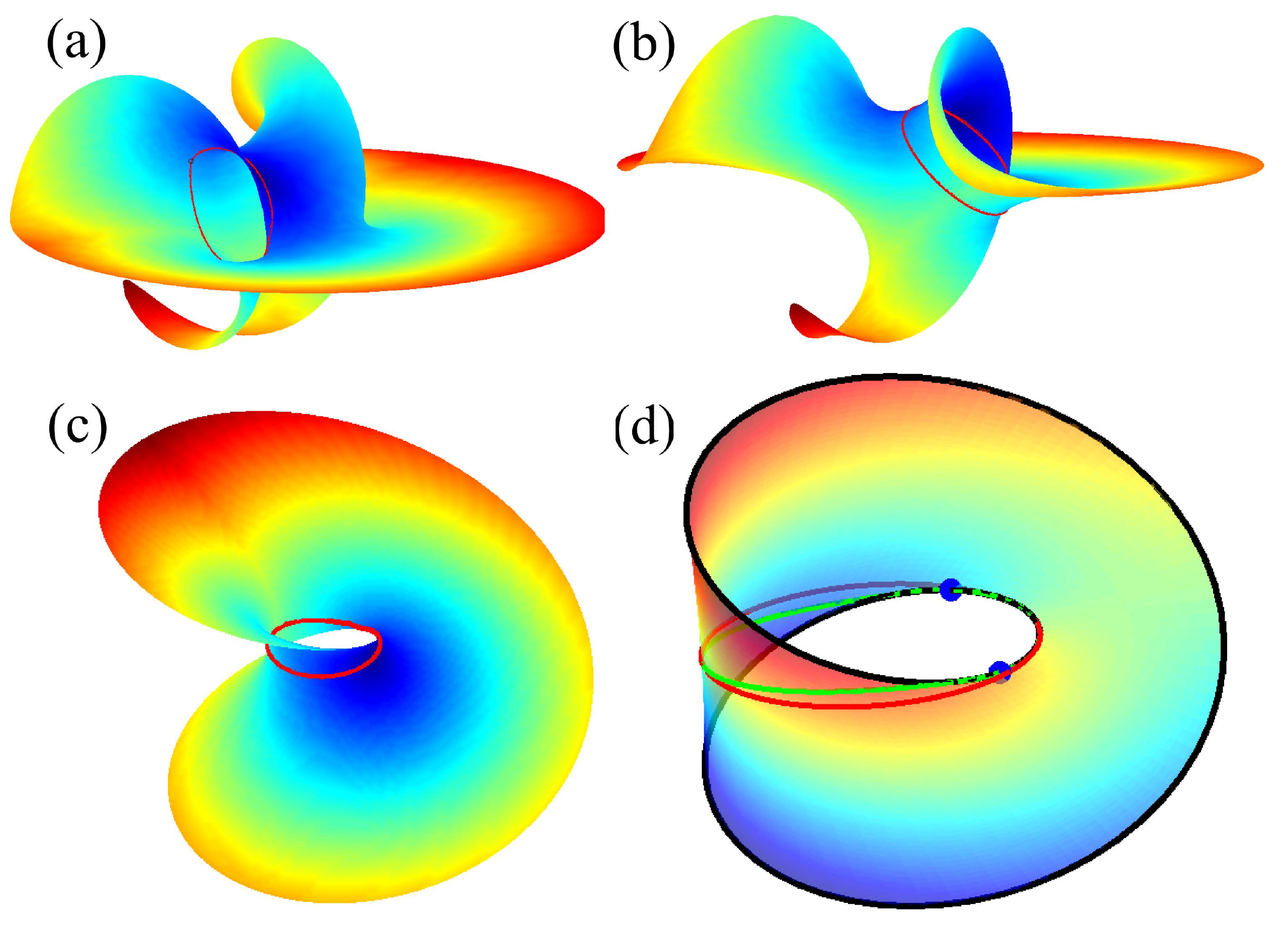}
\caption{Systoles on minimal surfaces.  (a,b) Two views of a punctured Klein bottle surface with a systole (red line). 
(c) Systole for a M{\"o}bius minimal surface. (d) Systole (green) and approximate systole (red) for a ruled 
surface M{\"o}bius surface.
Blue dots indicate loci where geodesic curvature changes sign.  Color scheme represents $z$-coordinate of points on
surfaces.  \label{fig5}}
\end{figure}

Figure \ref{fig5} shows the local systoles for stable minimal punctured Klein (a,b) and M{\"o}bius band (c) surfaces, where for the
latter the systole is unique.  
In the case of the punctured Klein bottle we  show only one systole - the second is a mirror image across the central vertical plane of symmetry.  
When the frame parameters are slightly beyond their critical values and the surfaces start 
their dynamic evolution towards the collapse the systoles of the stable configurations can be viewed as the initial  conditions for the
successive family of systoles described in the previous subsection.  A notable feature of each and every one of these systoles, 
including the one corresponding to the stable minimal surface,  is that they are topologically linked to the boundary.  In 
the case of the M{\"o}bius strip embedded in space, it can be shown that all non-contractible curves are in fact linked to the boundary.  
It is this linkage that 
provides a predictor of the location of the singularity.  In fact,  in the classic instability of the catenoid produced by pulling the two
rings apart, in which the systoles are never linked to the boundary, the singularities occur in the bulk. Moreover, when the 
catenoid collapse is forced by tilting the rings to induce a  quasi-static boundary singularity the systole eventually 
attaches itself to the boundary immediately before the surface reaches the singularity point; since this evolution has a stable minimal 
surface at all times the position of the systole prior to collapse also gives, even though trivially, an accurate prediction 
of the type  of singularity.

The reason systoles are important and useful in the prediction of the character of the motion may be understand through the
following heuristic argument:
The curvature associated with the systole is exactly canceled by 
its conjugate curvature when the surface is minimal. When the surface becomes unstable this delicate balance is broken and 
it  will move with a speed that varies from point to point, driven by the Laplace pressure difference, which is itself 
proportional to the mean curvature (note that both in mean curvature flow and in more realistic models it is this pressure
difference that determines the surface velocity). Since in the 
unstable regime the points with the largest mean curvature are generically those on the systole, they will move fastest and 
reach the singularity first by shrinking to a point. Furthermore, because of this shrinkage of the systole
and the impossibility of any point on the surface traversing the boundary,  the topological linkage of the systole and the 
boundary automatically determines the character (bulk or boundary) of the singularity.

Given that the M{\"o}bius band is the paradigm for singularity formation at the boundary, 
it follows that having an analytic
result for the systoles of this surface, even an approximate one, is  very desirable. With this  mind we have used the 
ruled surfaces \cite{Edge} generated by Eq. \ref{eqn1}, where now $-1 \le \mu \le 1$ \cite{GMPR}.
With the surface specified by the vector ${\bf x}(\mu,\theta)$, the length $L$ of a curve on that surface is
\begin{equation}
L=\int_0^{2\pi} d\theta \left[E\left(\frac{d\mu}{d\theta}\right)^2+2F\frac{d\mu}{d\theta}+G\right]^{1/2}
\end{equation}
where $E={\bf x}_{\mu}\cdot {\bf x}_{\mu}$, $F={\bf x}_{\mu}\cdot {\bf x}_{\theta}$, and 
$G={\bf x}_{\theta}\cdot {\bf x}_{\theta}$ are the coefficients of the first fundamental form.
To find the geodesic, we use a relaxation dynamics with $L$ as a Lyapunov function, 
\begin{equation}
\mu_t= -\frac{\delta L}{\delta \mu}~.
\end{equation}
Figure \ref{fig5}d shows the numerically obtained geodesic for the case $\tau=0.4$.
Observe that the solution hugs the boundary over a finite arc.  This is consistent with the general result that
a geodesic cannot touch the boundary if the geodesic curvature $\kappa_g$ of the boundary is positive, 
supported by the following argument. 
Suppose ${\bf \Gamma}(s)$ is a geodesic curve in a surface $\Sigma$,
parametrized by arc length $s$.  Then the second
derivative of ${\bf \Gamma}$ is normal to $\Sigma$, for otherwise, we could shrink
${\bf \Gamma}$ by pushing it in the direction of the second derivative.  If ${\bf \Gamma}$ touches the boundary at a convex point, 
then the second
derivative of ${\bf \Gamma}$ in the direction normal to the boundary but
tangent to $\Sigma$ must be positive, and we can shrink ${\bf \Gamma}$ by gently
pushing it in that direction.  It follows that moving a geodesic in the direction of the second derivative when the boundary curvature
is negative will push the curve off the surface. Hence, negative geodesic curvature of the boundary is 
a necessary but not sufficient condition for the systole to touch it.
In the present case, a straightforward calculation 
shows that 
\begin{equation}
\kappa_g=A(\tau)+B(\tau)\cos\theta+C(\tau)\cos 2\theta~,
\end{equation}
where $A,B,C$ are polynomials in $\tau$.  One can verify that the region of $\theta$ over which the numerical 
geodesic coincides with the boundary does indeed lie within the region of
negative $\kappa_g$.

A remarkably good analytic approximation to the systole on the ruled surface can be obtained by first finding a curve whose
tangent vector is everywhere orthogonal to both the surface normal and the lines of constant $\mu$.  Such a curve
does not lie on the surface. However, as the binormal to that curve is constant, it lies in a plane, which can be
easily found.  The approximate geodesic is then found as the intersection of the ruled surface and that plane.
This curve can be expressed explicitly in terms of its
Cartesian components or parametrically in the form $\mu(\theta;\tau)$ by first defining 
\begin{subequations}
\label{pollo}
\begin{eqnarray}
g(\tau)&=&\frac{2(1-\tau)}{\tau+2(1-\tau)^2}~,\\
\cos\theta^*&=&1/g~.
\end{eqnarray}
\end{subequations}
If Eq. \ref{pollo}b has real solutions for $\theta^*$ (which occurs for $0\le\tau\le 1/2$), then the geodesic is defined
piecewise:
\begin{equation}
\mu(\theta)=
\begin{cases}
-g(\tau)\cos\theta & \vert \theta\vert \ge\theta^*\\
-1 & {\rm otherwise}
\end{cases}
\end{equation}
Note that this ruled surface has the non-physical feature of self-intersection for $\tau>\tau_s\simeq 0.58$.  
For $1/2<\tau < \tau_s$ the approximate geodesic does not touch the boundary, and has the form $\mu(\theta)=-g\cos\theta$.
This is shown overlaid in red in Figure \ref{fig5}.  As with the numerically-obtained geodesic, the
region of coincidence with the boundary lies within the domain of negative geodesic boundary curvature, as required.

\section{Conclusions}

By identifying examples of boundary singularities, other than the collapse of the M{\"o}bius band, that occur when there is a 
topological change in soap films spanning a wire we have made a first attempt at classifying them. Besides the well-known
bulk singularity we found  two other types $ - $quasi-static and catastrophic$- $ both occurring at the boundary. 
We found  strong evidence that  any  catastrophic singularity which develops a neck will  asymptotically converge to a
M{\"o}bius like singularity. Furthermore, numerical studies showed that motion by mean curvature exhibits boundary singularities,
reproducing  the experimental findings.  We also found that the predictor of the position of the singularity was the linkage between
the boundary and the systole of the last stable surface before the onset of instability.

Clearly, these results are not in a rigorous framework,  but provide insight into which variables have a role in the
transition.  They also suggest that among the singularity types, the one corresponding to the 
quasi-static boundary should be the less difficult to study because it is the one for which it is possible to construct a family of
stable minimal surfaces at all times.  An important goal is to establish rigorously the conditions  under
which mean curvature motion leads to a  boundary singularity.  A more modest goal currently under study is to
prove, in the context
of the M{\"o}bius surface we have parametrized, that motion by mean curvature produces a singularity at the boundary.

\begin{acknowledgments}
We thank M. Gromov for comments on systoles
on bounded surfaces, L. Guth for suggesting the connection between
boundary curvature and systoles, R. Kusner and F. Morgan for discussions
on the topology of geodesics, and P. Constantin for discussions on mean
curvature flows; D. Page-Croft, C. Hitch, and J. Milton for technical assistance;
and K. Brakke, M. Evans, S. Furlan, and A. Kranik for computational
assistance. This work was 
supported by the 
Engineering and Physical Sciences Research Council (EPSRC) grant EP/I036060/1 and the Schlumberger Chair Fund. 
\end{acknowledgments}

\end{document}